# Interfacial Polytype Engineering of Polymer-Derived SiC via Compositionally Complex MXene Templating

Yuxiang Gan[1], Jianyu Dai[1], Laxmi Sai Viswanadha[2], Congjie Wei[1], Kelvin Y. Xie[3], Jeremy Watts[4], Mohammad Naraghi[2,3,5], Chenglin Wu[1*]

[1]Zachry Department of Civil & Environmental Engineering, Texas A&M University, College Station, Texas, 77843 USA

[2]Department of Mechanical Engineering, Texas A&M University, College Station, TX, 77843, USA

[3]Department of Materials Science and Engineering, Texas A&M University, College Station, TX, 77843, USA

[4]Department of Materials Science and Engineering, Missouri University of Science and Technology, Rolla, MO, 65409, USA

[5]Department of Aerospace Engineering, Texas A&M University, College Station, TX, 77843, USA

*Corresponding author, Email: chenglinwu@tamu.edu

## Abstract

Controlling polytype selection in polymer-derived silicon carbide (SiC) remains challenging since stacking sequences are determined locally at the nucleation front. Here, we demonstrate an interface-driven strategy to bias SiC polytype evolution by introducing compositionally complex TiVCrMoC$_3$ MXene nanosheets at the preceramic stage. Under spark plasma sintering (1900 °C, 70 MPa), which typically stabilizes cubic β-SiC, the MXene partially transforms into multicomponent carbide structures and generates two distinct heterogeneous interfacial states: reconstructed carbide/SiC interfaces that locally disrupt stacking sequences and promote hexagonal ordering, driving the emergence of α-SiC; and coherent MXene/SiC interfaces that preserve cubic stacking. Mechanical testing further reveals peak performance at an optimal MXene loading where interfacial reconstruction is most pronounced, with an ~82% increase in Young's modulus and ~42% improvement in fracture toughness. These findings highlight interfacial polytype engineering via two-dimensional carbide templates as a promising route for directing crystal structure evolution in polymer-derived ceramics.

Controlling the polytype of silicon carbide (SiC) remains a challenge in ceramic processing because different stacking sequences lead to distinct mechanical, thermal and electronic properties[1-3]. Although SiC can crystallize in more than 200 polytypes, polymer-derived SiC typically forms cubic β-SiC during high-temperature processing[4-7]. While processing parameters such as temperature and pressure can shift the conditions under which hexagonal α-SiC forms, they offer limited control over stacking sequence selection[8, 9]. This is because stacking sequences are established at the growth front during nucleation and early-stage growth, where local atomic environments influence how Si–C bilayers propagate[10]. One limitation of conventional ceramic composite processing is that filler-matrix interfaces form after crystallization has already occurred, limiting their ability to actively influence polytype selection[11, 12]. In contrast, polymer-derived ceramics (PDCs) allow secondary phases to be dispersed within the preceramic polymer prior to pyrolysis, establishing interfaces before SiC nucleation. These pre-existing interfaces enable embedded fillers to interact with the crystallization front during early-stage growth[13-15].

MXenes, a family of two-dimensional (2D) transition metal carbides and nitrides, are promising interfacial candidates due to their rich surface chemistry and structural tunability[16-18]. However, effective templating under spark plasma sintering (SPS) conditions requires that the interfacial phase retain structural and compositional coherence under extreme non-equilibrium environments where strong electrical and mechanical fields drive rapid densification. Under such conditions, structural transformation of MXenes is thermodynamically unavoidable. For conventional single-component MXenes such as $Ti_3C_2T_x$, high-temperature processing leads to termination loss, collapse of the layered structure, and transformation into TiC-like carbide phases. Compositionally complex MXenes offer a potential pathway. $TiVCrMoC_3$, derived from a multi-principal-element MAX phase precursor, forms a single-phase solid solution stabilized by configurational entropy[19].

Under high-temperature transformation, this entropy stabilization may promote reconstruction toward a compositionally homogeneous (Ti,V,Cr,Mo)$C_x$ carbide rather than a chemically segregated decomposition product, potentially preserving locally coherent interfacial regions with the growing SiC matrix. Such entropy-mediated reconstruction could enable MXene-derived interfaces to remain active during SiC crystallization. Combined with chemically active surface terminations (−O, −F, −OH) that interact with the preceramic network prior to nucleation[20-23], TiVCrMoC$_3$ is therefore proposed as a thermally resilient and compositionally stable interfacial template for directing SiC polytype evolution.

Here, we demonstrate that compositionally complex TiVCrMoC$_3$ MXene acts as an active interfacial template that biases the polytype evolution of polymer-derived SiC during SPS. Under conditions that typically stabilize cubic β-SiC (1900 °C, 70 MPa), the MXene undergoes localized interfacial reconstruction and partially transforms into a multicomponent carbide phase. This transformation generates spatially heterogeneous interfacial states, including reconstructed carbide/SiC interfaces and coherent MXene/SiC interfaces. The resulting interface-driven polytype modulation correlates with substantial improvements in mechanical performance at an optimal MXene loading. These findings establish this compositionally complex MXene as an active crystallization template and highlight two-dimensional carbide interfaces as a promising platform for interface-driven phase engineering in PDCs.

Figure 1 illustrates the interface templated crystallization pathway of MXene-SiC composites. Rather than being introduced after ceramic formation as in conventional powder mixed systems, the TiVCrMoC$_3$ MXene nanosheets are incorporated at the preceramic polymer stage, enabling the establishment of a dense population of MXene/polymer interfaces prior to SiC nucleation. This early-stage integration ensures molecular level contact between the MXene surface and the evolving Si-C network, positioning MXene as an active agent in the crystallization process rather than a passive reinforcement.

During subsequent spark plasma sintering at 1900 °C under 70 MPa, densification of the SiC matrix occurs simultaneously with structural transformation of the MXene. The non-equilibrium fields inherent to SPS drive localized interfacial reconstruction, leading to the partial conversion of TiVCrMoC$_3$ MXene into MXene-derived (Ti,V,Cr,Mo)C$_x$ phases embedded within the SiC matrix. This transformation generates spatially heterogeneous interfacial states, including reconstructed carbide/SiC interfaces and coherent MXene/SiC interfaces. The resulting interfacial heterogeneity defines the crystallization environment experienced by SiC during stacking formation. In this framework, reconstructed interfaces locally perturb stacking sequences, whereas coherent interfaces preserve cubic ordering, establishing the physical basis for interface-biased polytype evolution observed in subsequent structural analyses.

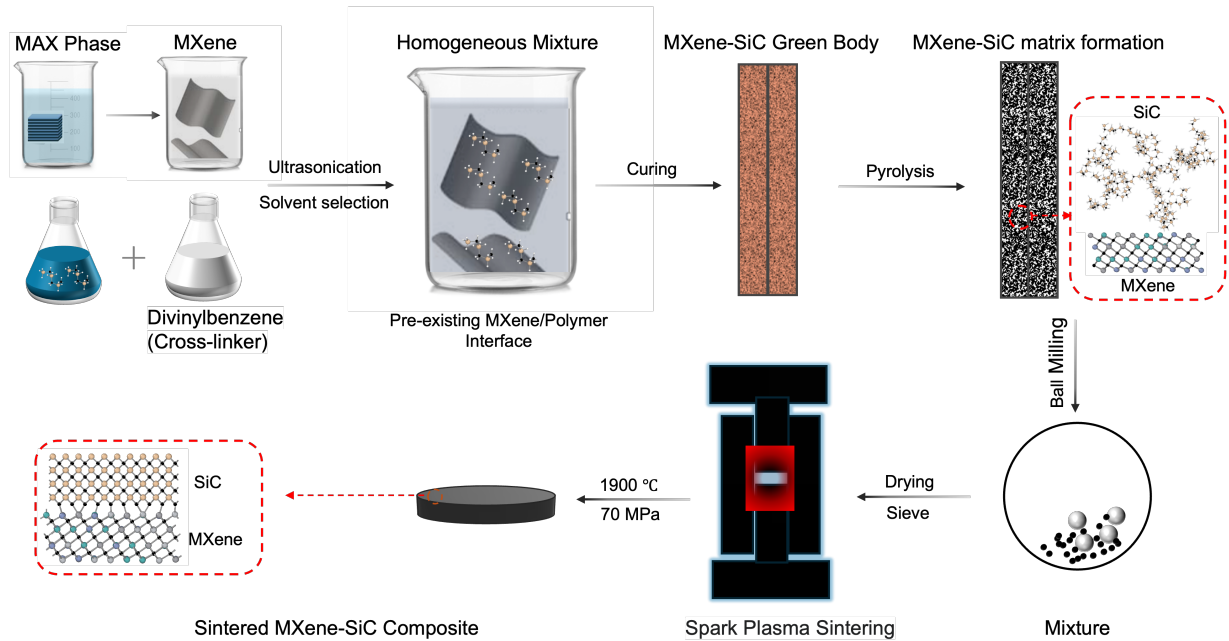

**Figure 1.** Schematic illustrating MXene-mediated SiC crystallization. MXene nanosheets are introduced at the preceramic stage, establishing pre-existing MXene/SiC interfaces prior to nucleation. During spark plasma sintering, MXene undergoes partial reconstruction into multicomponent carbide phases, generating heterogeneous interfacial states that locally influence SiC stacking evolution.

Establishing a well-defined MXene/SiC interfacial population is essential for isolating interface-driven crystallization effects. To achieve uniform MXene dispersion within the preceramic polymer, solvent-dependent processing was examined using five representative solvents (deionized (DI) water, ethanol, dimethyl sulfoxide (DMSO), tetrahydrofuran (THF), and N,N-dimethylformamide (DMF)). As shown in Figure 2a–b, the colloidal stability of MXene and its compatibility with the SMP-10 precursor vary significantly with solvent choice. Among the systems investigated, DMF enables the formation of a visually homogeneous MXene–polymer suspension without pronounced phase separation, whereas other solvents exhibit sedimentation, demixing, or instability[24]. SEM and EDS mapping (Figure 2c–g) further confirm that DMF processing produces the most uniform MXene distribution within the polymer-derived SiC matrix, with minimal restacking and aggregation. In contrast, samples prepared using other solvents display localized clustering. These results indicate that solvent selection primarily governs the spatial distribution of MXene and therefore the density of pre-existing interfaces prior to crystallization. DMF processing is therefore adopted to establish a uniform interfacial landscape for probing interface-driven polytype evolution in the following sections.

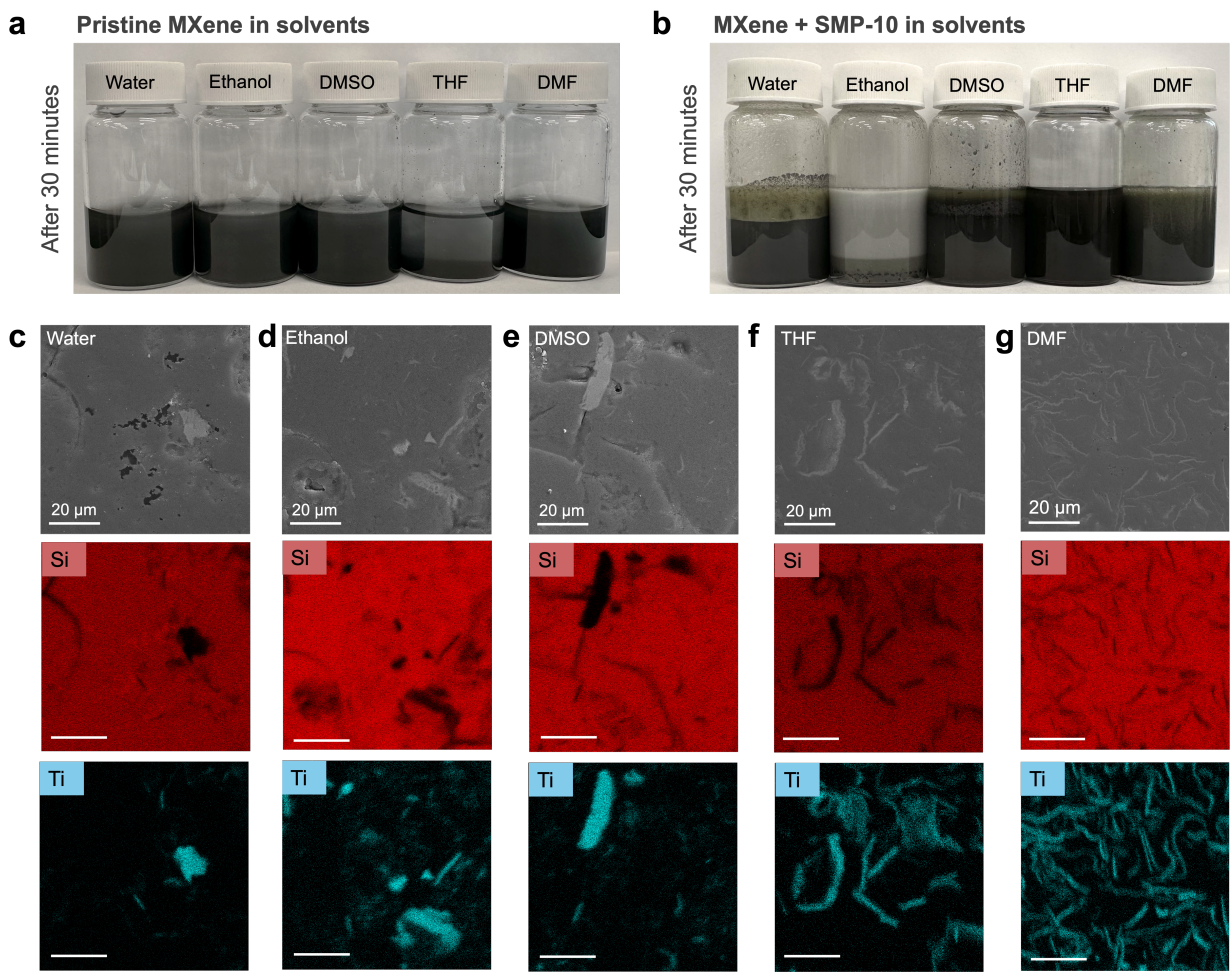

**Figure 2**. Solvent-dependent MXene dispersion and resulting interfacial population in polymer-derived SiC composites. (a) Colloidal stability of MXene in different solvents. (b) Compatibility of MXene–SMP-10 mixtures. (c–g) SEM and EDS mapping showing solvent-controlled MXene distribution, highlighting

uniform interfacial populations achieved with DMF processing (Si and Ti signals represent the distributions of the SiC matrix and MXene, respectively).

XRD provides the first evidence of interface-biased polytype evolution in MXene–SiC composites (Figure 3a). The pristine SMP-10 derived ceramic exhibits strong diffraction peaks indexed to cubic 3C-SiC (β-SiC), corresponding to the (111), (200), (220), and (311) planes, together with minor graphite reflections[25]. The 3C-SiC phase adopts a cubic zinc blende structure characterized by an ABC stacking sequence of Si–C bilayers as shown in Figure 3 (c, left), which is the typical product of polymer-derived SiC processing.

With the incorporation of TiVCrMoC$_3$ MXene, new diffraction peaks (2θ) appear around 33°, and 39°, corresponding to hexagonal 6H-SiC (α-SiC) along with weak reflections of a MXene derived (Ti,V,Cr,Mo)C$_x$ phase. The 6H-SiC polytype shows an ABCACB stacking sequence along [0001] direction, as shown in Figure 3 (c, right). The formation of the 6H SiC phase becomes more pronounced at MXene loadings of 3 wt% and above. The coexistence of both β-type and α-type reflections in the composite indicates that the β to α polytypic transformation occurs during the sintering process[26, 27].

This observation is particularly significant because the processing conditions employed (1900 °C, 70 MPa) are reported to favor stabilization of the cubic 3C-SiC phase and suppress the β to α transformation during spark plasma sintering[28]. This is confirmed with the observation of SMP-10 results shown in Figure 3a, where control samples sintered under identical conditions without MXene exhibit predominantly β-SiC. Therefore, the appearance of α-SiC reflections only in the MXene-containing composites indicates the influence of MXene interfaces in the polytype selection of SiC, rather than a phenomenon solely affected by the processing conditions.

Although the two polytypes possess fundamentally different stacking arrangements (Figure 3c), their close-packed planes 3C-SiC (111) and 6H-SiC (0006) have identical interplanar spacing, causing their strongest reflections to overlap at the same 2θ position. Polytype difference instead arises from non-close-packed reflections, such as the (10-11) and (10-13) planes of 6H-SiC (Supporting Information, Figure S6), which have sharper peaks and higher intensities at higher TiVCrMoC$_3$ MXene concentrations. To further support peak assignment, DFT-derived reference patterns are included in Figure 3d, e, with additional computational details and reference structures provided in the Supporting Information (Figures S12–S16). The simulated diffraction fingerprints highlight the intrinsic structural and diffraction differences between 3C-SiC and 6H-SiC, serving as comparisons to aid interpretation of the experimental XRD data, particularly in regions where close-packed reflections overlap.

This spatially heterogeneous phase evolution is intrinsic to the SPS process, which generates non-uniform thermal and mechanical fields. Local temperature gradients, pressure variations, and stress concentrations create regions that experience different thermodynamic driving forces for phase transformation[29, 30]. Consequently, regions experiencing sufficiently high local temperatures and stresses undergo the β-SiC to α-SiC transformation and facilitate the partially conversion of compositionally complex TiVCrMoC$_3$ MXene into the (Ti,V,Cr,Mo)C$_x$ phase, while cooler or lower-stress regions retain β-SiC or partially preserved MXene structures[31].

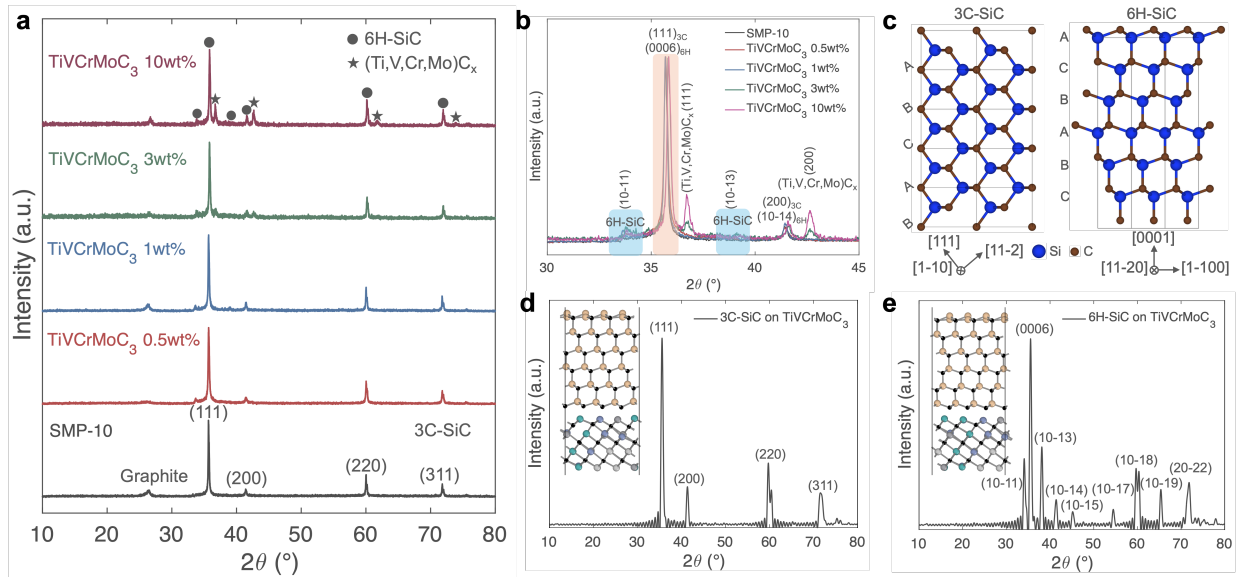

**Figure 3.** (a) XRD patterns of the spark plasma sintered MXene-SiC composites with different TiVCrMoC$_3$ MXene contents (0.5 wt%–10 wt%). (b) Enlarged diffractograms in the 2θ = 30–45° range highlighting the emergence of 6H-SiC reflections and (Ti,V,Cr,Mo)C$_x$ phases with increasing MXene addition. (c) Structural models of 3C-SiC and 6H-SiC illustrating their ABC and ABCACB stacking sequences along the close-packed directions. (d, e) Simulated radial distribution functions and corresponding reference diffraction features for 3C-SiC and 6H-SiC, included as diagnostic comparisons to assist phase identification in the experimental XRD patterns.

TEM observations directly reveal the interfacial origin of SiC polytype modulation in MXene–SiC composites (Figure 4). Low-magnification TEM identifies MXene-derived carbide platelets embedded within the SiC matrix, consistent with the heterogeneous phase evolution inferred from XRD. High-resolution TEM (HRTEM) at reconstructed interface regions (Figure 4b) reveals SiC lattice fringes with d-spacing of 2.52 Å, corresponding to the close-packed planes shared by 3C-SiC (111) and 6H-SiC (0006). FFT analysis indicates that the local stacking cannot be indexed to a single polytype, revealing stacking disorder with mixed cubic–hexagonal sequences (inserted images in Figure 4b-c). This provides strong evidence that reconstructed carbide/SiC interfaces locally perturb SiC stacking, embedding short-range hexagonal sequences within a predominantly cubic framework. Imaging within the MXene-derived phase (Figure 4c) further shows lattice spacings consistent with an FCC (Ti,V,Cr,Mo)C$_x$ carbide, with measured spacings of $d_{(220)}$ = 1.50 Å and $d_{(311)}$ = 1.27 Å. These values agree with the theoretical (220) and (311) interplanar spacings of an FCC carbide having lattice parameters consistent with the XRD results. The corresponding FFT pattern, acquired along the [112] zone axis, displays a set of diffraction spots indexed to the (111), (220), and (311) planes. This microstructural evidence is also supported by previous studies reporting the collapse of MXene layers and their transformation into carbide phases at elevated temperatures[32, 33]. The corresponding high-angle annular dark-field scanning transmission electron microscopy (HAADF-STEM) image and EDS mappings (Supporting Information, Figure S8) confirm uniform elemental distribution across the (Ti,V,Cr,Mo)C$_x$ phase region, with clear localization of Ti, V, Cr, and Mo signals.

Reconstructed interfaces are not the only interfacial state. In additional regions (Figure 4d–f), the MXene/SiC interface remains coherent. The SiC (111) planes are aligned parallel to the MXene basal plane and FFT analysis confirms preservation of cubic 3C ordering. This crystallographic alignment is further illustrated by the corresponding atomic models (Figure 4f), indicating that the interface is geometrically compatible and preserves cubic stacking without undergoing interfacial reconstruction. Additional representative coherent interfaces are presented in the Supporting Information (Figure S10).

The coexistence of reconstructed and coherent interfaces therefore points to a threshold-driven interfacial reconstruction regime. Within this framework, reconstructed carbide interfaces promote local stacking rearrangement, whereas coherent MXene/SiC interfaces preserve the conventional cubic stacking pathway. Electron localization function (ELF) calculations confirm that electron localization within the compositionally complex TiVCrMoC$_3$ interfacial layer remains electronically consistent across both configurations, while polytype-dependent differences are localized within the SiC region (Supporting Information, Figure S15). It provides computational validation of the interface-directed crystallization mechanism.

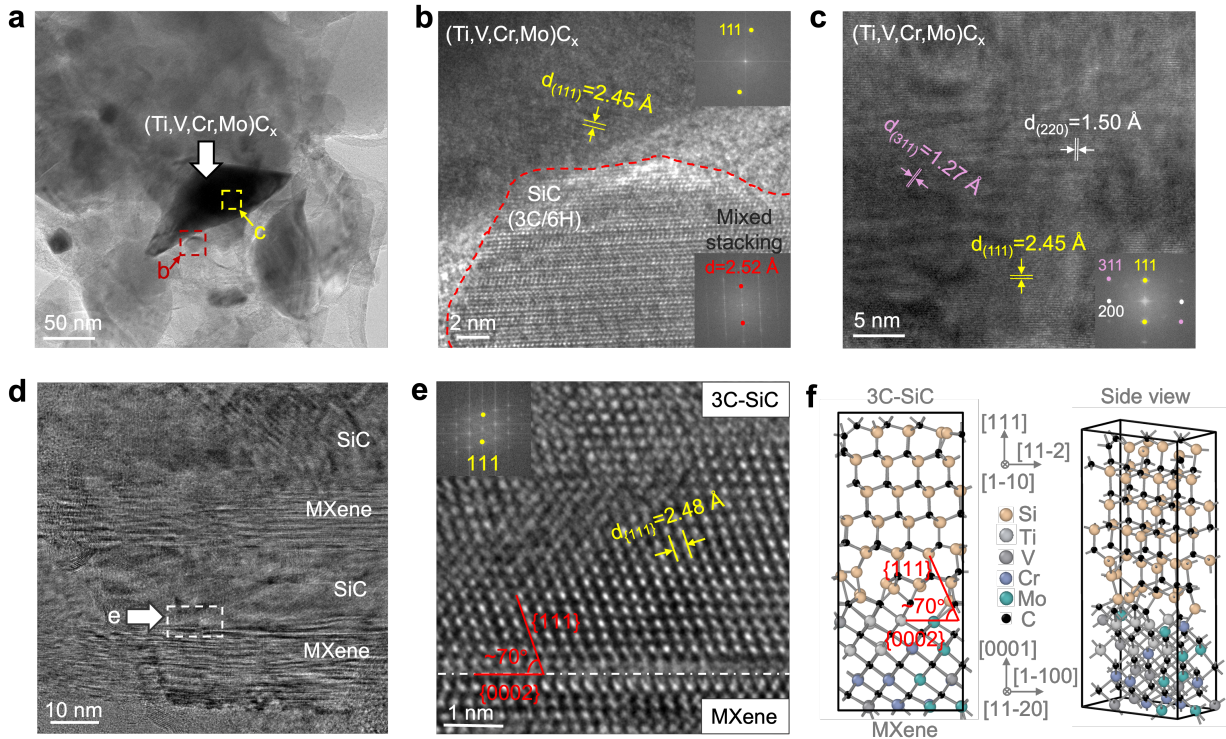

**Figure 4.** Atomic-scale evidence of interface-biased SiC crystallization in MXene–SiC composites after SPS. (a) Low-magnification TEM showing an embedded MXene-derived (Ti,V,Cr,Mo)C$_x$ platelet within the SiC matrix. (b) HRTEM of a reconstructed carbide/SiC interface exhibiting mixed cubic–hexagonal stacking signatures in adjacent SiC. (c) HRTEM from the carbide region confirming an FCC (Ti,V,Cr,Mo)C$_x$ phase. (d–f) Coherent MXene/SiC interface: (d) HRTEM of an intact MXene/SiC interface, (e) zoom-in showing 3C-SiC (111) planes parallel to the MXene basal surface with FFT-confirmed cubic ordering, and (f) atomic models illustrating geometric registry at the coherent interface.

Figure 5 summarizes the mechanical response of MXene–SiC composites as a function of TiVCrMoC$_3$ MXene content in both pyrolyzed and SPS-consolidated states. In the pyrolyzed specimens (Figure 5a, c), MXene incorporation leads to systematic increases in Young's modulus, hardness, and fracture toughness, with all properties reaching a maximum at an optimal loading of 3 wt%. This trend indicates that mechanical behavior is governed by interfacial state rather than MXene content alone. At this composition, the elastic modulus increases by ~55% compared to the pristine SMP-10 derived ceramic. The fracture toughness exhibits a substantial increase of ~49% compared to the unreinforced matrix, highlighting a marked improvement in damage tolerance in addition to stiffness. Further increasing the MXene content beyond 3 wt% results in a deterioration of mechanical performance, as observed for the 10 wt% composite. SEM and EDS analyses reveal localized MXene agglomeration and interfacial heterogeneity at high loading (Supporting Information, Figure S4), which are likely to limit effective reinforcement.

Following SPS consolidation (Figure 5b, d), all samples exhibit substantially higher mechanical properties due to matrix densification. The dependence on MXene content remains consistent with the pyrolyzed case: the 3 wt% TiVCrMoC$_3$ composite again displays the highest Young's modulus, hardness, and fracture toughness. At this optimal loading, the elastic modulus reaches approximately 346 GPa, representing an ~82% improvement over the pristine sintered SMP-10 matrix. In parallel, the fracture toughness increases by ~42%, demonstrating that the toughness enhancement is retained after ultra-high temperature processing. Increasing MXene content beyond this regime leads to reduced performances despite higher filler concentration, consistent with interfacial heterogeneity and MXene aggregation limiting effective load transfer.

These trends suggest that increasing interfacial population raises the probability of reconstructed interfaces, which govern load transfer and crack deflection. The simultaneous enhancement of stiffness and toughness is therefore attributed to interface-mediated mechanisms, including efficient load transfer across MXene-derived carbide/SiC interfaces and crack deflection at interfacial regions (Figure 5e–f). The composition corresponding to peak mechanical performance coincides with the regime where MXene transformation and SiC stacking reconstruction are most pronounced, providing macroscopic validation of the interface-biased crystallization mechanism established above.

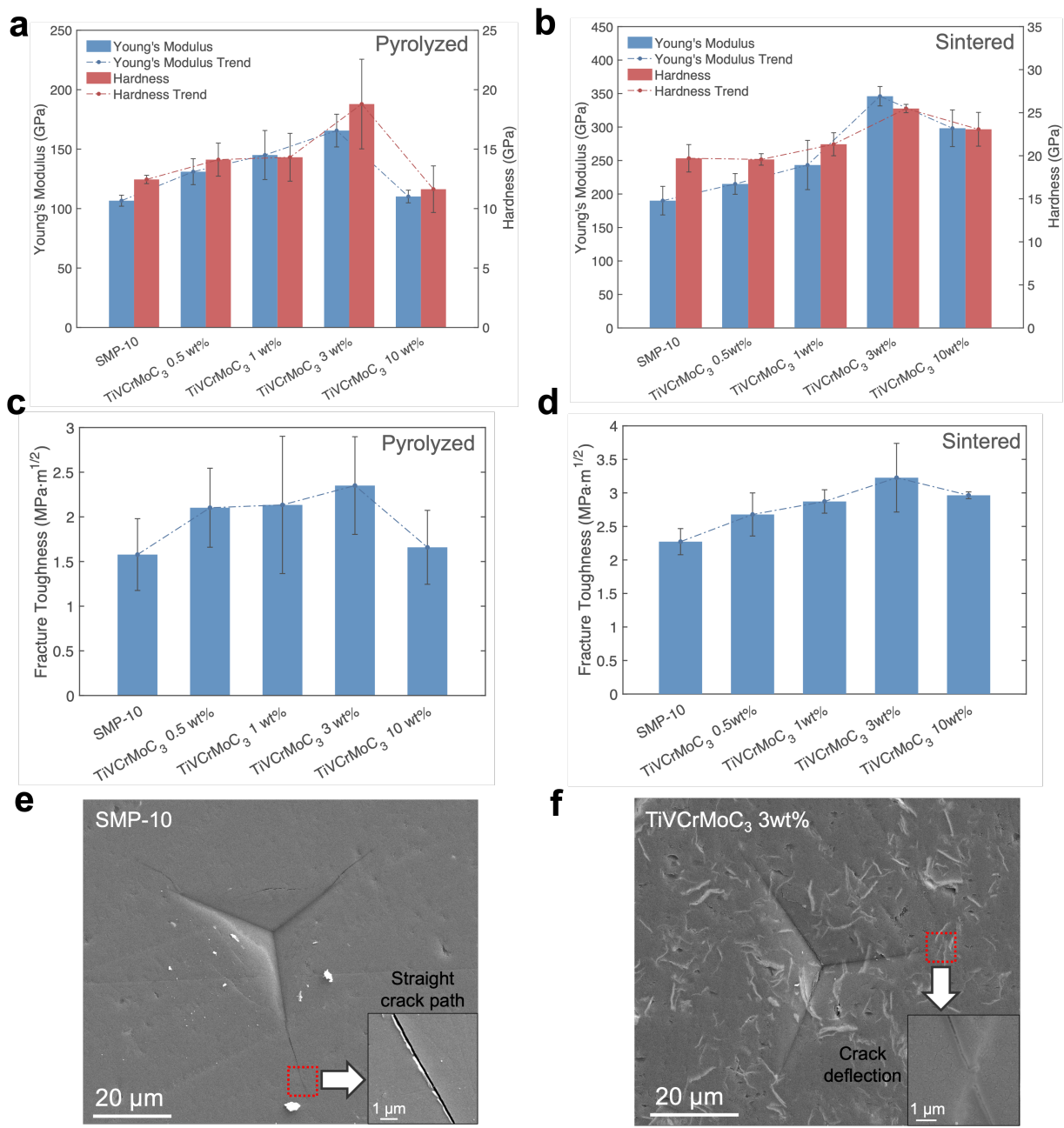

**Figure 5.** Mechanical response of MXene–SiC composites as a function of TiVCrMoC$_3$ MXene content. (a) Young's modulus and hardness of pyrolyzed MXene–SiC composites with varying MXene loadings. (b) Young's modulus and hardness of SPS-consolidated MXene–SiC composites with varying MXene loadings. (c) Fracture toughness of pyrolyzed MXene–SiC composites as a function of MXene content. (d) Fracture toughness of SPS-consolidated MXene–SiC composites as a function of MXene content. (e) Pristine SMP-

10 derived SiC ceramic showing straight radial cracks extending from the indent corners, indicative of brittle fracture. (f) MXene-containing composite showing shortened and deflected cracks; inset highlights crack deflection at interfacial regions.

In summary, this work demonstrates that compositionally complex TiVCrMoC$_3$ MXene acts as a thermally resilient and active interfacial template that biases the polytype evolution of polymer-derived SiC during SPS. Under processing conditions that typically stabilize cubic β-SiC, MXene undergoes localized interfacial reconstruction and partially transforms into a multicomponent FCC carbide, generating spatially heterogeneous interfacial states. Reconstructed carbide/SiC interfaces locally perturb stacking sequences and promote hexagonal ordering within a predominantly cubic matrix, while coherent MXene/SiC interfaces preserve cubic stacking. The resulting composites exhibit simultaneous enhancements in stiffness and fracture toughness at the optimal MXene loading, with an ~82% increase in Young's modulus and fracture toughness improvements exceeding 40%. These findings establish interface-templated crystallization as a viable strategy for directing phase selection and highlight two-dimensional carbide templates as a general platform for interfacial phase engineering in PDCs.

## Acknowledgments

The authors gratefully acknowledge the support of the US Office of Naval Research under Grant number N00014-23-1-2009. Additionally, they acknowledge the use of Materials Characterization Facility (RRID: SCR_022202).

## Author Contributions

Yuxiang Gan synthesized the MXene–ceramic composites, performed the experiments and materials characterization, and drafted and revised the manuscript. Jianyu Dai conducted DFT and

MD simulations. Laxmi Sai Viswanadha assisted with experiments and provided technical input. Congjie Wei reviewed the manuscript. Kelvin Y. Xie assisted with the TEM characterization, reviewed the manuscript. Jeremy Watts and Mohammad Naraghi supervised the project, contributed to conceptualization, and secured funding. Chenglin Wu supervised and guided the experimental work, contributed to conceptualization and theoretical modeling, reviewed the manuscript, and provided funding support. All authors discussed the results and approved the final version of the manuscript.

**Notes**

The authors declare no competing financial interest.